\documentclass[prl,showpacs,twocolumn,floatfix,superscriptaddress]{revtex4}
\usepackage{graphicx}
\usepackage{amsmath,amssymb,mathrsfs}

\newcommand\varH{\mathscr{H}}%
\newcommand{\beq}{\begin{equation}}
\newcommand{\eeq}{\end{equation}}
\newcommand{\bea}{\begin{eqnarray}}
\newcommand{\eea}{\end{eqnarray}}

\begin{document}

\title{RKKY and magnetic field interactions in coupled Kondo quantum dots}

\author{Pascal Simon}
\affiliation{Laboratoire de Physique et Mod\'elisation des Milieux
  Condens\'es, CNRS and UJF, 38042 Grenoble, France}
\author{Rosa L\'opez}
\affiliation{D\'epartement de Physique Th\'eorique,
  Universit\'e de Gen\`eve, CH-1211 Gen\`eve 4, Switzerland}
\author{Yuval Oreg}
\affiliation{Department of Condensed Matter Physics, Weizmann Institute of Science, Rehovot, 76100 Israel}

\begin{abstract}
We investigate theoretically the transport properties of two independent artificial Kondo impurities. They are coupled together via a tunable Ruderman-Kittel-Kasuya-Yoshida (RKKY) interaction. For strong enough antiferromagnetic RKKY interaction, the impurity density of states increases with the applied in-plane magnetic field. This effect can be used to distinguish between antiferromagnetic and ferromagnetic RKKY interactions. These results may be relevant to explain some features of recent experiments by Craig et al. [cond-mat 0404213].
\end{abstract}

\maketitle

\emph{Introduction.---}Local exchange interactions between itinerant electrons 
 and localized magnetic impurities in diluted alloys results in the Kondo effect. Its main signature
is the formation of a very narrow peak at the
 Fermi level ($E_F$) in the density of states (DOS) of
the localized spin at low enough temperatures $T\ll T_K$, where $T_K$ is the Kondo temperature \cite{hew93}.  Magnetic impurities, even for tiny concentrations,
also interact with one another via the Ruderman-Kittel-Kasuya-Yoshida (RKKY) interaction,
an indirect spin-spin interaction mediated by conduction electrons. 
The competition between both interactions 
leads to extremely rich and various behaviors such as spin glasses, non Fermi liquid
fixed points, etc.~\cite{hew93} Part of this complexity and richness is already embodied in the
\emph{two-impurity Kondo problem}~\cite{jayaprakash,Jones,Affleck,Krishna-murthy80a}. 

Progressive advance in manufacturing nanostructures has enabled to build up 
quantum dots that can act as a single spin-1/2~\cite{dot}. 
This has motivated the emergence of an exciting area of research based on the spin 
degree of freedom, termed \emph{spintronics}, with promising applications in quantum 
computing~\cite{loss}. 
Experiments have been able to observe the Kondo effect in quantum dots~\cite{experiment}, its most remarkable signature being the  
\emph{ unitary limit} of the linear conductance at
zero temperature \cite{theory}. This is a consequence of the formation of a singlet state 
between the delocalized electrons and the localized spin in the dot.
Continuous improvements of the fabrication techniques in semiconductor nanostructures
have made possible the realization of more complicated structures like  
double quantum dots. They can mimic the behavior of two artificial magnetic impurities coupled by
a direct tunnel coupling \cite{blick}. In some recent experiments Craig ~{\it et al.} \cite{craig} have investigated a device in which two quantum dots are connected to a common open conducting region. Here both the RKKY and Kondo interactions compete, thus providing an \emph{experimental realization of the two-impurity Kondo problem}. 
These authors found that the strength of the  non-local RKKY interaction can be used to
control the Kondo effect in one dot by tuning gate parameters of the other dot.
This constitutes the first observation of the RKKY indirect interaction    
in quantum dots opening the road toward the realization
of multiple artificial impurity configurations with a high degree of 
tunability and a non-local spin control.

\begin{figure}\centering
\includegraphics*[width=85mm]{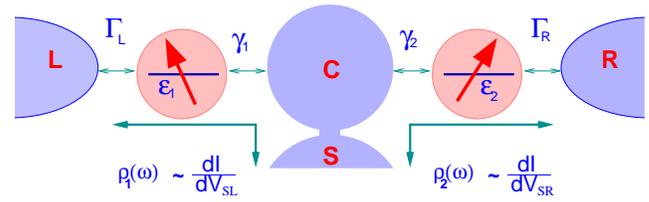}
\caption{(Color online). Schematic picture of the setup containing two quantum dots. Each dot $1(2)$ is strongly connected to an open central conducting region C through the tunneling couplings $\gamma_{1(2)}$ and very weakly connected to its respective lead $L(R)$ by $\Gamma_{L(R)}$.
 Thus, the differential conductance ($dI/dV_{\rm SL(R)}$) between the $S$ and $L(R)$ through the dot 1(2) will reflect the DOS in dot 1(2).}
\label{scheme}
\end{figure}

In this Letter we study the geometry  depicted in Fig.~\ref{scheme} that corresponds to the experimental situation in [\onlinecite{craig}]. Two small quantum dots are strongly connected to a central larger open area. The whole system can be regarded as a triple dot. Two extra (left and right) leads are also weakly connected to the triple dot. These are used
as transport arms to probe individually the DOS in each dot without perturbing the whole triple dot structure much like a STM tip acts when probing the DOS at the top of a magnetic impurity.
A priori, three energy scales are present in the system at zero temperature: the Kondo temperatures of each individual dots ($T_K^1$, $T_K^2$) and the strength of the RKKY interaction $I$. When an in-plane magnetic field is applied we need to consider the Zeeman energy as well.
Here we analyze the transport properties in the triple dot system for various parameter ranges in order to reproduce the main features
of the experiment [\onlinecite{craig}]. We also predict that an in-plane applied magnetic field
competes with the antiferromagnetic (AF) RKKY interaction and may induce a new Kondo effect in a way reminiscent of what happens in  
quantum dots with an integer number of electrons \cite{magnetic,nazarov}. 
 
\emph{Model.---}
In the geometry depicted in Fig.~\ref{scheme} the two quantum dots are two magnetic impurities of spin-1/2 strongly coupled to the middle region (C) via hybridization widths $\gamma_1,\gamma_2$. The dots are weakly connected to the leads through $\Gamma_L,\Gamma_R$ such that, $\Gamma_{L(R)}\ll\gamma_{1(2)}$. A simple Hamiltonian able to describe the system without the leads reads:
\begin{multline}
\label{hkondo}
\varH =\varH_C
+ J_1\vec{S}_1\cdot\vec{\sigma}_1 + J_2\vec{S}_2\cdot
\vec{\sigma}_2+ I(\phi)\vec{S}_1\cdot\vec{S}_2 \,.
\end{multline}
In Eq.~(\ref{hkondo}) $\varH_C$ describes a grain with a continuous DOS. $\vec{S}_{1(2)}$ is the spin operator of the $1(2)$ dot and $\vec{\sigma}_{1(2)}$ is the local spin DOS in region C coupled to the $1(2)$ dot. The Kondo coupling of dot $1(2)$ is given by $\pi\rho_cJ_{1(2)}\approx\gamma_{1(2)}\left\{\left[1/\left(-\varepsilon_{1(2)}\right)\right]+\left[1/\left(U_{1(2)}+\varepsilon_{1(2)}\right)\right]\right\}$ where $\varepsilon_{1(2)}$ is the position of the last occupied level  controlled by the dot gate voltage, $U_{1(2)}$ is the intradot Coulomb interaction and $\rho_c$ is the density of states in the middle region C. The RKKY interaction is $I(\phi)=J_1 J_2 \frac{\vartheta}{D} {\rm cos(\phi)}$  where $D$ is the bandwidth of the conducting region, $\vartheta$ is a constant with depends on the geometry of the middle conducting region and $\phi$ controls the sign of the RKKY interaction~\cite{noteRKKY}. The Kondo energy scales for each dot are defined by $T_K^{1(2)}\sim D\exp{(-1/2\rho_c J_{1(2)})}$.

\emph{Ferromagnetic RKKY interaction.---}
Let us start with $I(\phi)<0$. Depending on the relative strength of the
ferromagnetic interaction and the Kondo scales one finds different physical
scenarios. For a \emph{small} ferromagnetic coupling, either $|I|\ll T_K^1, T_K^2$ or $T_K^2 < |I| <T_K^1$, it is easy to show using renormalization group (RG) arguments that both dot spins are screened independently and each dot DOS shows the usual Kondo resonance. However, for a \emph{large} ferromagnetic coupling, ($|I|\gg T_K^1, T_K^2$) the spins of the dots add up in a $S=1$ state.  This state is screened in a two-stage 
procedure by conducting channels formed by the even and odd linear combination of the electrons in the central conducting region~\cite{jayaprakash}. This defines two different Kondo temperatures, $T_K^{\rm even}$ and $T_K^{\rm odd}$ that are much lower than $T_K$ ($=T_K^1=T_K^2$ for $J_1=J_2$)~\cite{jayaprakash}. When $T$ is lowered until  $T\sim{\rm min}(T_K^{\rm even},T_K^{\rm odd})$ $1/2$ unit of the $S=1$ state is efficiently screened while the other half is free. By further lowering the temperature, $T\ll {\rm min}(T_K^{\rm even},T_K^{\rm odd})$, the $S=1$ state is completely screened leading to a Fermi liquid behavior. Here, the linear conductance between the source and the left lead is $\mathcal{G}_{(\rm {SL})}\sim 2e^2/h [4\Gamma_L\gamma_1/\left(\Gamma_L+\gamma_1\right)^2]$.
To summarize, for both cases, 
namely, \emph{small} and \emph{large} $I$ compared with the Kondo scales and at very low temperatures ($T\rightarrow 0$) the DOS of each dot displays a Kondo resonance.

In the measured nonlinear conductance~\cite{craig}, the zero bias anomaly (ZBA) disappears when increasing $I$ and two small peaks at source-left lead voltage $eV_{\rm SL}/2\sim \pm I$ show up at equal
distance from $E_F=0$ corresponding either to singlet-triplet ($I>0$) or to triplet-singlet ($I<0$) states. A priori the sign of $I$ is
not known.
This suggest two possibilities (i) the RKKY-like interaction is
antiferromagnetic in the experiment or (ii) is ferromagnetic with an experimental temperature $T>{\rm max}(T_K^{\rm even},T_K^{\rm odd})$
preventing therefore the formation of the ZBA mentioned above. 
We shall demonstrate below how, by using an in-plane magnetic field, we can
distinguish between both situations. Our results show a positive sign for the RKKY-like interaction observed in the experiment.

\emph{Antiferromagnetic RKKY interaction.---}
Let us now  provide a quantitative analysis of the case $I>0$.
The competition between the Kondo effect and the AF ordering results in a quantum critical point when $(I/T_K)_c\simeq 2.54$ in situations where there is a symmetry between even and odd parity channels \cite{Jones} ($T_K$ is the same Kondo temperature for both magnetic impurities). When this symmetry is broken the critical transition is replaced by a crossover~\cite{Jones}. When $I/T_K>(I/T_K)_c$ we have the AF phase where the two impurities are combined into a singlet state. However when the Kondo coupling becomes stronger, $I/T_K<(I/T_K)_c$, each impurity prefers to form its respective Kondo singlet with the delocalized electrons. This is the Kondo phase. Now the question is whether this competition is manifested in the transport properties in our geometry where we allow to have  different Kondo temperatures for the $1$ and $2$ dots (e.g., with $\varepsilon_1\neq\varepsilon_2$ or $\gamma_1\neq\gamma_2$).
 
Our results are based on a strong-coupling approximation, the
so-called slave-boson mean field theory (SBMFT), which is known to
capture the main physics in the Kondo regime~\cite{sbmft}. The SBMFT has been recently applied to study the Kondo effect in
double quantum dots~\cite{georges,eto,ros02}. In
particular, in Ref.~[\onlinecite{ros02}] it was shown that the
nonlinear conductance ${\mathcal G}(V)\equiv dI/dV$ for serial and parallel double quantum dots, directly reflects the physics of the competition between the Kondo effect and the superexchange interaction.

\begin{figure}\centering
\includegraphics*[width=85mm,angle=0]{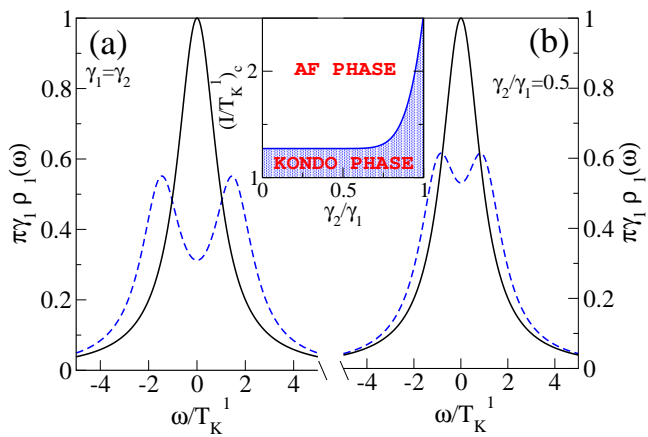}
\caption{(Color online). (a) DOS for dot 1 in the \emph{symmetric} case ($\gamma_1=\gamma_2$)  for $I/T_K^1=1$ (solid line) and  $I/T_K^1=3$ (dashed line). The critical value at which the splitting occurs is $(I/T_K^1)_c\approx 2.54.$ (b) DOS for dot 1 in the \emph{asymmetric} case ($\gamma_2=\gamma_1/2$) for $I/T_K^1=1$ (solid line) and  $I/T_K^1=2$ (dashed line). Here the critical value at which the splitting occurs is $(I/T_K^1)_c \approx 1.27$.
Inset: Critical value $(I/T_K^1)_c$ vs $\gamma_2/\gamma_1$ at which the transition from the Kondo state to a singlet state occurs. The boundary (denoted by the solid line) separates both phases.}
\label{fig2}
\end{figure}

\emph{Results.---}
The critical value at which the transition from the Kondo state (KS) to the AF phase takes place can be obtained by comparing their ground state energies~\cite{eto}: $E_{GS}^{K}-E_{GS}^{AF}=I/4-(2/\pi) T_K$ (assuming $T_K=T_K^1=T_K^2$). Thus, the transition can be estimated to occur at $(I/T_K)_c=8/\pi\simeq 2.54$. However, in our geometry this critical value changes because both dots are not equally coupled to the central region.  We find that the critical point depends now on both Kondo temperatures as follows~\cite{eto}: 
\begin{equation}
\left(\frac{I}{T_K^{1}}\right)_c=\frac{4}{\pi}\left(1+\frac{T^2_{K}}{T_{K}^1}\right)\,.
\end{equation}
For $0\leq T^2_{K}\leq T_{K}^1$ we have $4/\pi \leq (I/T_K^{1})_c\leq 8/\pi$. Since $T^1_{K}$, and $T^2_{K}$ depend exponentially on the tunneling couplings and the level positions, a small asymmetry between these parameters induces a huge change in the ratio $T^2_{K}/T_{K}^1$.
 In order to illustrate this effect we plot in the inset of Fig.~\ref{fig2} the KS$\rightarrow$AF transition as a function of the asymmetry between the tunneling couplings $\gamma_2/\gamma_1$. Already with a slight asymmetry (around $10\%$) the transition lowers down to $(I/T_K^{1})_c=(4/\pi)\approx 1.27$ in good agreement with the experimental value $(I/T_K^1)_c=1.2$. 

We solve the set of mean field equations within the SBMFT~\cite{georges,eto,ros02} for $\varepsilon_{1}=-3.5$ and $D=60$.
This leads to a Kondo temperature for the dot 1: $T_K^1=10^{-3}\gamma_1$. All energies are given in units of $\gamma_1=1$. We  focus on two cases: (i) the symmetric case with $\gamma_1=\gamma_2$ corresponding
to $T_K^1=T_K^2$ [Fig. \ref{fig2}(a)]
and (ii) the strongly asymmetric case $\gamma_2=\gamma_1/2$ with $T_K^2\ll T_K^1$  [Fig. \ref{fig2}(b)]. Both plots are alike except that the splitting of the Kondo peak occurs at different values of $I/T_K^1$ since the KS$\rightarrow$AF transition takes place at different critical values. When $I/T_K^{1}< (I/T_K^{1})_c$, the RKKY interaction is not strong enough to link both spins into an AF singlet state and the DOS displays a Kondo resonance.
For the asymmetric case, the value of $I/T_K^{1}$ is increased above the critical value
upon reducing $\varepsilon_1$ but keeping always $T_K^2$ very small compared to $T_K^1$. In both cases the DOS changes dramatically when increasing $I$ above the critical value showing a double peak structure with peaks at $\omega/T_K^1=\pm I/2$. 
These results clearly shows that measurement of the splitting yields an estimate of the intensity of the RKKY interaction.

\emph{Effect of a magnetic field.---}
In what follows we focus on the symmetric situation, i.e., $T_K^1=T_K^2$. A local magnetic field 
$H$  is applied in the plane of the dots. It couples equally to the spins of both dots giving rise to an additional Zeeman term $-\vec H(\vec S_1+\vec S_2)$ in Eq. (\ref{hkondo}). When $I=0$, a strong magnetic field destroys
the Kondo effect and a splitting of the DOS occurs for $T_K\lesssim H$. For $|I|< T_K$, the spins of the dots are almost independently screened and the magnetic field ($H> T_K$) leads to a splitting of the ZBA. 
When the RKKY interaction is ferromagnetic with $|I|>T_K$ the two spins are locked into a $S=1$ state and at sufficiently low temperature the DOS displays a Kondo resonance. By applying a magnetic field $|I|\gg H>{\rm max}(T_K^{\rm even},T_K^{\rm odd})$ the DOS becomes split. As a result the conductance versus $H$ always decreases as in the usual $S=1/2$ Kondo effect.

As we anticipated when $I>0$ the interplay between the magnetic field and the RKKY interaction results in very different transport regimes. Next, we analyze in detail this case by using the SBMFT~\cite{ros02,note2}.
\begin{figure}\centering
\includegraphics*[width=85mm]{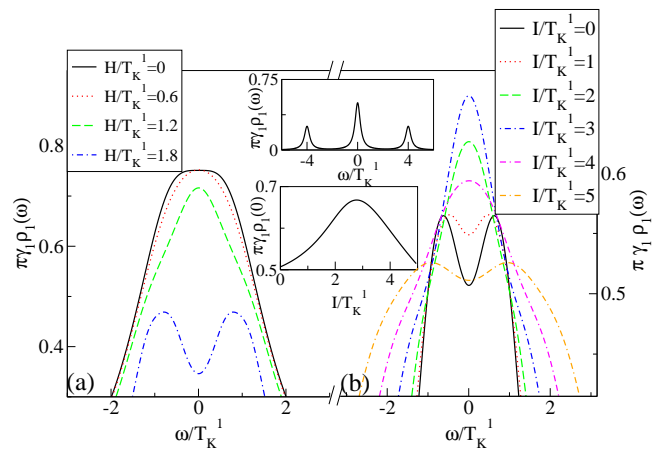}
\caption{(Color online). (a) DOS for the dot 1 for $I=1.5T_K^1$ 
at different values of the magnetic field $H$.(b) DOS for the dot 1 for $H=1.5T_K^1$ at different values of $I/T_K^1$.
Upper inset: DOS for the dot 1 for $H=I=4T_K^1$. 
Lower inset: DOS at $E_F$ for the dot 1 as a function of $I/T_K^1$ for $H=1.5T_K^1$. All the curves correspond to the symmetric case $\gamma_1=\gamma_2$ and $T_K^1=T_K^2$.}
\label{fig3}
\end{figure}
 Figure~\ref{fig3} summarizes our results. In the left panel, we fix $I/T_K^1<(I/T_K^1)_c$ increasing the magnetic field $H$ [Fig.~\ref{fig3}(a)]. Striking enough, we need a much higher value of the magnetic field ($H\approx 1.8 T_K$) to split the DOS when the RKKY interaction is present. Clearly, both the RKKY and the Zeeman interactions compete. Whereas 
$H$ tries to align both spins, the AF RKKY interaction tries to
 align antiferromagnetically the spins of the dots. 
This corresponds to the experimental result where such a behavior of
 the nonlinear conductance with the magnetic field has been reported, suggesting that $I>0$. 
In order to corroborate this picture we have considered the reversed situation
in Fig.~\ref{fig3}(b): we fix the magnetic field $H=1.5T_K^1$ increasing $I$.  In the absence of the RKKY interaction, the DOS displays a double peak structure with peaks located at $\pm H/2$~\cite{notemag}. By increasing $I$ one clearly observes that the splitting of the DOS is destroyed and a resonance at $E_F$ emerges, restoring the Kondo effect around some critical magnetic field $H_{c}$. By further increasing $I$ the Kondo resonance splits again. 
In general for $H,I\gg T_K^1$, we find four peaks in the DOS at $\pm|H\pm I|/2$ corresponding to the possible ground states of the two antiferromagnetically coupled $S=1/2$ impurities in a magnetic field~\cite{regularization}. Here, the restoration of the KS occurs at the critical value $H_c\approx I$ where two peaks merge together and give a central Kondo resonance as it shown in the upper inset of Fig.~\ref{fig3}.
The value $H_c$ corresponds to the special point where the AF singlet state and the lowest component of the triplet state become almost degenerate because of the presence of $H$~\cite{note3}.  This gives rise to a new magnetic-field induced Kondo effect rather similar to the one predicted in a quantum dot with an even number of electrons~\cite{magnetic,nazarov}. Nevertheless, it is worth pointing out that this occurs here as a consequence of the \emph{non-local RKKY coupling}. This new magnetic-field induced Kondo effect is governed by a new Kondo scale $T_K(H_c)$ which is of order $T_K$ in our SBMFT analysis.
$T_K(H_c)$ can be evaluated using the poor man's scaling treatment developed in Ref.~[\onlinecite{nazarov}],
$T_K(H_c)/T_K\sim (T_K/H)^\alpha$ where $\alpha\geq\pi/2-1$ depends on 
the ratio between the local DOS of conduction electrons in the even and odd (under parity) sectors 
and is minimum when this ratio is 1.  
Note also that the dot DOS at $E_F$ shows a nonmonotonous behavior under the influence of an in-plane magnetic field because of the antiferromagnetic nature of the RKKY interaction (see lower inset in Fig.~\ref{fig3}). The measurement of a nonmonotonous behavior of the conductance versus $I/T_K^1$ would provide an unambiguous way of determining the sign of the RKKY coupling. 

\emph{Conclusions.---} 
We have shown that different transport regimes arise in the geometry studied experimentally in [\onlinecite{craig}] depending on the sign of the RKKY-like interaction $I$. We find that these
different regimes can be distinguished when a (strong) in-plane magnetic field is applied. In particular, the antiferromagnetic RKKY interaction competes with 
the effect of a magnetic field and from both interactions a magnetic-field induced Kondo effect emerges leading to an enhancement of the conductance with the magnetic field. 

We acknowledge M. B\"uttiker and D. S\'anchez for helpful discussions. This research was supported by the EU RTN No. HPRN-CT-2000-00144 and the Spanish MECD (RL), by Minerva and by DIP grant c-7.1 (YO).

\emph{Note added.---} During the completion of this work we became aware of related work by M.G~Vavilov and L.I.~Glazman~\cite{vavilov} which investigates complementary aspects on this problem.


\begin{thebibliography}{90}
\bibitem{hew93} A.C.~Hewson, {\it The Kondo Problem to Heavy Fermions}
  (Cambridge University Press, Cambridge, UK, 1993).
\bibitem{jayaprakash}
C.~Jayaprakash, H.R.~Krishna-murthy and J.W.~Wilkins, Phys. Rev. Lett. {\bf 47}, 737 (1981).
\bibitem{Jones}
B.A.~Jones, C.M.~Varma, Phys. Rev. Lett. \textbf{58}, 843
  (1987); B.A.~Jones, C.M.~Varma, and J.W. Wilkins {\it ibib.}
\textbf{61}, 125 (1988). B.A. Jones and C.M. Varma, Phys. Rev. B \textbf{40}, 324 (1989).
\bibitem{Affleck}
I.~Affleck and A.W.W.~Ludwig, Phys. Rev. Lett. \textbf{68}, 1046 (1992). I.~Affleck, A.W.W.~Ludwig, and B.A.~Jones, Phys. Rev. B \textbf{52}, 9528
  (1995).
\bibitem{Krishna-murthy80a} H.R.~Krishna-murthy, J.W.~Wilkins, and
  K.G.~Wilson, Phys. Rev. B \textbf{21}, 1003 (1980).
\bibitem{dot} For a review, see L.P.~Kouwenhoven {\it et al}., in {\it
    Mesoscopic Electron Transport}, edited by L.L.~Sohn {\it et al}
    (Kluwer, Dordrecht, 1997).
\bibitem{loss} D. Loss and D. DiVicenzo, Phys. Rev. Lett. A {\bf 57}, 120 (1998
\bibitem{experiment} D. Goldhaber-Gordon {\it et al}., Nature {\bf 391}, 156
  (1998). S.M.~Cronenwett {\it et al}.,
  Science {\bf 281}, 540 (1998). J.~Schmid {\it et al}.,
  Physica B {\bf 256-258}, 182 (1998).
\bibitem{theory} L.I.~Glazman and M.E.~Raikh, Pis'ma Zh. \'Eksp. Teor. Fiz.
  {\bf 47}, 378 (1988) [JETP Lett. {\bf 47}, 452 (1988)];
  T.K.~Ng and P.A.~Lee, Phys. Rev. Lett. {\bf 61}, 1768 (1988).
\bibitem{blick} A.W.~Holleitner {\it et al}., Phys. Rev. Lett. {\bf 87}, 256802 (2001).
\bibitem{craig} N.J.~Craig {\it et al}., cond-mat/0404213 (2004).
\bibitem{magnetic} M.~Pustilnik {\it et al}., Phys. Rev. Lett. {\bf 84}, 1756 (2000); D. Giuliano and A. Tagliacozzo, {\it ibid.} {\bf 84}, 4677 (2000).
\bibitem{nazarov} M.~Eto and Y.~Nazarov, Phys. Rev. B {\bf 64}, 085322 (2001). M.~Pustilnik and L.I.~Glazman, {\it ibid.} {\bf 64} 045328 (2001).
\bibitem{noteRKKY} 
$\phi$ controls the sign of the RKKY interaction and  depends on the Fermi wavelength and on geometrical parameters like the separation between the two dots.
\bibitem{sbmft} P.~Coleman, Phys. Rev. B {\bf 29}, 3035 (1984). For a review,
  see D.M.~Newns and N.~Read, Adv. Phys. {\bf 36}, 799 (1987).
\bibitem{georges}
A.~Georges and Y.~Meir, Phys. Rev. Lett. {\bf 82}, 3508 (1999).
\bibitem{eto}
T.~Aono and M.~Eto, Phys. Rev. B {\bf 63}, 125327 (2001).
\bibitem{ros02}
R.~Lopez, R.~Aguado, and G.~Platero Phys. Rev. Lett. {\bf 89}, 136802 (2002).
\bibitem{note2} A very small direct tunneling term between the dots has been added to allow a continuous transition between the AF singlet and Kondo state in the SBMFT approximation (see Refs.~[\onlinecite{georges,ros02}]). 
\bibitem{notemag} The SBMFT is known not to correctly predict the peak position which are located at $\pm \alpha H$ with $2/3<\alpha=\alpha(H/T_K)<1$ [see J.E.~Moore and X.-G.~Wen, Phys. Rev. Lett. {\bf 85}, 1722 (2000)].
\bibitem{regularization}
For large $H$, fluctuations are important and the mean field solution gives unphysical vanishing renormalized tunneling couplings leading to divergences in the dot propagators that we regularize by hand. Going beyond mean field would regularize these divergences. 
\bibitem{note3} In the KS$\rightarrow$AF transition the DOS evolves smoothly from a single peak (KS) to a double peak structure located at $\pm\kappa$  with $0\leq\kappa\leq I/2$.
Here $\kappa$ characterizes the strength of the AF order between the impurities (see Ref.~[\onlinecite{georges,ros02}]). Thus for $I\approx T_K^1$ the needed $H_c$ to restore the Kondo effect is $H_{c}\approx 2 \kappa\leq I$ (right panel in Fig 3). For large $I\gg T_K^1$, the AF state is completely established and the DOS shows the usual peaks at $\pm I/2$ ($\kappa$ is maximum). In this case one finds as expected $H_c\sim I$ for the critical value of the magnetic field able to restore the Kondo peak (see upper inset in Fig 3).
\bibitem{vavilov} M.G.~Vavilov and L.I.~Glazman, cond-mat/0404366.
\end{thebibliography}
\end{document}